\documentclass[11pt]{article}

\usepackage{amssymb,amsmath,amsfonts,amsthm,graphicx}

\def\be{\begin{equation}}
\def\ee{\end{equation}}
\def\bea{\begin{eqnarray}}
\def\eea{\end{eqnarray}}

\def\1{\'{\i}}                           

\def\m{{\eta}}  

\def\k{\omega}
\def\>#1{{\bf #1}}

 \def\adsw{AdS$_\omega$\ }

\def\tfrac#1#2{ {\scriptstyle { \frac {#1}{#2}}}}
\def\pois#1#2{\left\{ {#1},{#2} \right\}}
\def\conm#1#2{\left [ {#1},{#2} \right ]}

\def\C{{\Upsilon}}
\newcommand\vt{\vartheta}

\begin{document}

\thispagestyle{empty}

\
\medskip
\medskip

 \vskip1cm

 \begin{center}

\noindent {\Large{\bf {Quantum groups and noncommutative spacetimes\\[4pt]  with cosmological constant}}}\\

\end{center}

\medskip
\medskip
\medskip

\noindent {\sc A.~Ballesteros$^\dagger$\footnote{Based on the contribution presented at  ``Eighth International Workshop DICE2016",   
 Castello Pasquini in Castiglioncello, Italy, September 12-16, 2016.}, I.~Guti\'errez-Sagredo$^\dagger$, F.J.~Herranz$^\dagger$, C.~Meusburger$^\ddagger$ and P.~Naranjo$^\dagger$}

\medskip

{\small
\noindent
$^\dagger$Departamento de F\1sica, Universidad de Burgos, Plaza Misael Ba\~nuelos s.n., E-09001 Burgos, Spain

\noindent
$^\ddagger$Department Mathematik,  FAU Erlangen-N\"urnberg, Cauerstr.~11, D-91058 Erlangen, Germany

\medskip

\noindent E-mail:  {angelb@ubu.es, igsagredo@ubu.es, fjherranz@ubu.es, catherine.meusburger@math.uni-erlangen.de, pnaranjo@ubu.es}
}

\medskip

\begin{abstract}  
\noindent
Noncommutative spacetimes are widely believed to model some properties of the quantum structure of spacetime at the Planck regime. In this contribution the construction of (anti)de Sitter noncommutative spacetimes obtained through quantum groups is reviewed. In this approach  the quantum deformation parameter $z$ is related to a Planck scale,  and the cosmological constant $\Lambda$ plays the role of a second deformation parameter of geometric nature, whose limit $\Lambda\to 0$ provides the corresponding noncommutative Minkowski spacetimes. 
\end{abstract}


\section{Introduction}

Non-Abelian algebras play a prominent role in the Hamiltonian description of physical systems. For instance, nonrelativistic Quantum Mechanics is based on  a `noncommutative phase space' in which position and momenta operators generate the Lie algebra 
\be
\conm{\hat x_i}{\hat p_j}=i\hbar\,\delta_{ij},
\qquad
\conm{\hat x_i}{\hat x_j}=0,
\qquad
\conm{\hat p_i}{\hat p_j}=0,
\qquad
i,j=1,\dots, N,
\label{quantum}
\ee
which is the direct sum of $N$ copies of the Heisenberg--Weyl algebra. Here, noncommutativity is controlled by the fundamental constant $\hbar$, since the $\hbar\to 0$ limit of~\eqref{quantum} leads to an Abelian algebra, and the classical limit of~\eqref{quantum} is defined as
$
\pois{x_i}{p_j}:=\lim_{\hbar\to 0}{\frac{\conm{\hat x_i}{\hat p_j}}{i \hbar}}=\delta_{ij}.
$
In this way we recover the symplectic structure of Classical Hamiltonian Mechanics, which can be properly said to be a Poisson-noncommutative theory. 

On the other hand,  spacetimes with nonvanishing cosmological constant $\Lambda$ lead to noncommuting momenta in a natural way. For instance,  
the (3+1) (anti-)de Sitter (hereafter (A)dS)  Lie  algebra of isometries of the corresponding spacetimes can be written as
\be
\begin{array}{lll}
[J_i,J_j]=\epsilon_{ijk}J_k ,& \quad[J_i,P_j]=\epsilon_{ijk}P_k  ,&\quad
[J_i,K_j]=\epsilon_{ijk}K_k ,\\[2pt]\displaystyle{
[P_i,P_j]=-\k\,\epsilon_{ijk}J_k},&\quad\displaystyle{[P_i,K_j]=-\,\delta_{ij}P_0}
,    &\quad\displaystyle{[K_i,K_j]=-\,\epsilon_{ijk} J_k}
,\\[6pt][P_0,P_i]=\k  K_i ,&\quad [P_0,K_i]=-P_i  ,&\quad[P_0,J_i]=0 ,
\end{array}
\label{3mas1}
\ee
where $i,j= 1,2,3$, $\epsilon_{123}=1$ and $\k=-\Lambda$. Here  $\{J_i,P_0,P_i, K_i\}$ denote  the generators of rotations, time translation, space translations and boosts. The algebra~\eqref{3mas1} is called the  \adsw Lie algebra. In this framework, when $\k>0$ we recover the (3+1) anti-de Sitter Lie algebra $so(3,2)$, when $\k<0$ we find the (3+1) de Sitter  Lie algebra $so(4,1)$, and the flat limit $\k\to 0$ gives rise to the (3+1) Poincar\'e   algebra $iso(3,1)$. Note that the commutation rules amongst the   translation generators read 
$$
[P_i,P_j]=-\k\,\varepsilon_{ijk} J_k,
\qquad
[P_0,P_i]=\k  K_i,
$$
and the fundamental parameter that controls the noncommutativity of momenta is just the cosmological constant  $\k=-\Lambda$.

In this contribution we shall deal with a third type of noncommutativity: The one arising in different approaches to Quantum Gravity aiming to describe the `quantum' structure of the geometry of spacetime at the Planck scale through a noncommutative algebra
of `quantum spacetime coordinates'~\cite{Snyder, Woronowicz, FredenCMP, Szabo, Verlinde}. In this approach the Planck length $l_P$ (or energy $E_p$) is the parameter that governs the noncommutativity of the spacetime algebra, thus generating uncertainty relations between noncommuting coordinates that can be used in order to describe a  `fuzzy' or `discrete' nature of the spacetime at very small distances or high energies~\cite{Garay, Maggiore}. 

It is worth recalling that most of the `quantum' spacetimes that have been introduced so far are noncommutative versions of the Minkowski spacetime~\cite{kMas, kMR, kZakr,kappaP,nullplane}.  As a consequence, the construction of noncommutative spacetimes with nonvanishing cosmological constant arises as a challenging problem in order to describe the interplay between  the nonvanishing curvature of spacetime and quantum gravity effects,  having in mind the possible cosmological consequences of Planck scale physics~\cite{BHBruno, Starodutsev, Marciano, iceCUBE, Giulia1,Giulia2}.

In this paper we report on recent results~\cite{BHMplb, BHMCQG, BHMNsigma, BHMNplb, BHNplb, BHMNkappa, BMNphs} concerning the construction of noncommutative (A)dS spacetimes, which have been obtained by making use of the theory of quantum groups. We recall that the so-called `quantum' deformations of Lie groups and algebras (see~\cite{KR,Dri87, Jimbo, Takh, FRT, CP, majid} and references therein) present many features that make them suitable to be considered in a Quantum Gravity scenario:

\begin{itemize}

\item They are Hopf algebra deformations of kinematical Lie groups in which the quantum deformation parameter, hereafter  $z= \ln q$, can be related to a Planck scale parameter. 

\item They give rise to noncommutative spacetimes which are covariant under quantum group (co)actions. In this context, several  notions in `quantum kinematical geometry' can be rigorously generalized, like Poisson and quantum homogeneous spaces~\cite{DrHS, Koor, Zakrzewski, Reyman, Ciccoliqplanes}.

\item Quantum groups can be thought of as Hopf algebra quantizations of Poisson--Lie (PL) groups, and the relevance of the latter in (2+1)   gravity has been rigorously established (see~\cite{AMII, FR, AT, Witten1, cm1, cm2, bernd1} as well as the connections between dynamical $r$-matrices and gauge-fixing~\cite{cmt}).

\item  Deformed Casimir operators can be interpreted as modified dispersion relations of the same type that appear in several phenomenological approaches to Quantum Gravity~\cite{AEN, Amelinodispersion1, Mattingly}.

\end{itemize}

Indeed, quantum group techniques can only be  used to construct  noncommutative analogues of spacetimes that can be obtained either as group manifolds or as homogeneous spaces, but Minkowski and (A)dS  spacetimes fall into this class. Moreover, each Lie group admits a number of inequivalent quantum group deformations, and the quantum spacetime arising from each of them can be essentially different. The classification and explicit construction of such plurality of quantum geometries constitutes one of the main issues in the theory of quantum kinematical groups, which is far from being completed. In general, the results contained in the next sections show that noncommutative spacetimes with cosmological constant can be viewed as `geometric' nonlinear deformations (with parameter $\Lambda$) of the noncommutative Minkowski spacetimes with quantum deformation parameter $z$ related to either $l_P$ or $E_p$. 

The paper is organized as follows. In the next section we summarize the construction of noncommutative spacetimes from quantum groups. Section 3 is devoted to present two different noncommutative (2+1) (A)dS spacetimes that have been obtained by making use of two quantum deformations whose classical counterpart is known to be compatible with the Chern-Simons approach to (2+1) gravity, together with a brief summary of the results obtained so far in (3+1) dimensions.
 

\section{Quantum groups, Lie bialgebras and noncommutative spacetimes}

Quantum groups can be understood as noncommutative generalizations of algebraic groups, and all their properties are mathematically encoded within the Hopf algebra of functions on the  group. Alternatively, they can also be interpreted as quantizations of  PL  groups, {\em i.e.}, of the Poisson-Hopf algebras of multiplicative Poisson structures on Lie groups. It is well known that PL structures on a (connected and simply connected) Lie group $G$ are in one-to-one correspondence with Lie bialgebra structures $(g,\delta)$ on $g=\mbox{Lie}(G)$~\cite{DriPL}, where the skewsymmetric cocommutator map $
\delta:{g}\to {g}\wedge {g}
$
fulfils the two following conditions:
\begin{itemize}
\item i) $\delta$ is a 1-cocycle, {\em  i.e.},
\be
\delta([X,Y])=[\delta(X),\,  Y\otimes 1+ 1\otimes Y] + 
[ X\otimes 1+1\otimes X,\, \delta(Y)] ,\qquad \forall \,X,Y\in
{g}.
\label{1cocycle}\nonumber
\ee
\item ii) The dual map $\delta^\ast:{g}^\ast\wedge {g}^\ast \to
{g}^\ast$ is a Lie bracket on ${g}^\ast$.
\end{itemize}
Therefore, each quantum group $G_z$ (with quantum deformation parameter $z= \ln q$) can be put in correspondence with a PL group $G$, and the latter with a Lie bialgebra structure $(g,\delta)$.

Moreover, in the same manner as Lie algebras provide the infinitesimal version of Lie groups around the identity, quantum algebras play the same role with respect to quantum groups.  
more explicitly, quantum algebras $U_z(g)$ are Hopf algebra deformations of universal enveloping algebras $U(g)$, and are constructed as formal power series in a deformation parameter $z$
and coefficients in $U(g)$. The Hopf algebra structure in $U_z(g)$ is provided by a coassociative coproduct map
$
\Delta_z: U_z(g)\longrightarrow U_z(g)\otimes U_z(g)
$,
which is an algebra homomorphism, together with its associated counit $\epsilon$ and antipode $\gamma$ mappings. It can be easily shown that  the first-order deformation (in $z$) of the coproduct map
\be
\Delta_z
=\Delta_0 + z\,\delta+ o[z^2] ,
\label{powerco}\nonumber
\ee
is just the Lie bialgebra cocommutator map $\delta$, where we have denoted the primitive (nondeformed) coproduct for $U(g)$ as $\Delta_0(X)=X \otimes 1+1\otimes X $. Again, each quantum deformation is related to a unique Lie bialgebra structure $(g,\delta)$. 
If we consider a basis for $g$ where
\be
[X_i,X_j]=C^k_{ij}X_k ,
\label{liealg}\nonumber
\ee
any cocommutator $\delta$ will be of the form
\be
\delta(X_i)=f^{jk}_i\,X_j\wedge X_k \, ,
\label{precoco}\nonumber
\ee
where $f^{jk}_i$ is the structure tensor of the dual Lie algebra $g^\ast$, that will be given by
\be
[\hat\xi^j,\hat\xi^k]=f^{jk}_i\,\hat\xi^i \, ,
\label{dualL}
\ee
where $\langle  \hat\xi^j,X_k \rangle=\delta_k^j$. In particular, if $G$ is a group of isometries of a given spacetime (for instance,  (A)dS  or Poincar\'e), then $X_i$ will be the Lie algebra generators and $\hat\xi^j$ will be the local coordinates on the group. We thus realize that if the cocommutator $\delta$ is non-vanishing (which means that we have a non-trivial deformation of $U(g)$) then, automatically, the commutator~\eqref{dualL} among the spacetime coordinates associated to the translation generators of the group will be non-zero. This is just  the way in which noncommutative spacetimes arise from quantum groups. Moreover, we stress that, as we will see in what follows, higher-order contributions to the noncommutative spacetime are obtained from dualizing the full quantum coproduct  $\Delta_z$. 

In some cases the 1-cocycle $\delta$ is found to be    coboundary 
\be
\delta(X)=[ X \otimes 1+1\otimes X ,\,  r],\qquad 
\forall\,X\in {g} ,
\label{cocom}
\ee
where $r$  (the classical $r$-matrix)
$
r=r^{ab}\,X_a \wedge X_b\, ,
$
has
to be a solution of the modified classical Yang--Baxter equation  
\be
[X\otimes 1\otimes 1 + 1\otimes X\otimes 1 +
1\otimes 1\otimes X,[[r,r]]\, ]=0, \qquad \forall X\in {g},
\label{mCYBE}\nonumber
\ee
where the Schouten bracket is defined as
$
[[r,r]]:=[r_{12},r_{13}]+ [r_{12},r_{23}]+ [r_{13},r_{23}] 
$, 
  where $r_{12}=r^{ab}\,X_a \otimes X_b\otimes 1, \, r_{13}=r^{ab}\,X_a \otimes 1\otimes X_b, \, r_{23}=r^{ab}\,1 \otimes X_a\otimes X_b$ (recall that $[[r,r]]=0$ is just the classical Yang--Baxter equation). An important application of the classical $r$-matrix associated to a given quantum deformation consists in the fact that the PL structure on $G$ linked to $\delta$ is explicitly given by the so-called Sklyanin bracket
\be
\{f,g\}=  r^{ij}(X^L_i \,f\, X^L_j\, g -
X^R_i\, f\, X^R_j\, g) , \label{gb}\nonumber
\ee
where  $X_i^L$, $X_i^R$ denote the right- and left-invariant vector fields on $G$.  The linearization of this bracket in terms of the local coordinates in $G$ is just the Poisson version of~\eqref{dualL}, and the quantization of the Sklyanin bracket gives the commutation relations that define the quantum group $G_z$ in terms of noncommuting local coordinates. Moreover, if we compute the Sklyanin bracket for the spacetime coordinates we will obtain the 
`semiclassical' version of the noncommutative spacetime associated to the quantum group $G_z$, and the main algebraic features induced from the spacetime noncommutativity can be straightforwardly appreciated at this Poisson level, such as,  for instance, the role played by the cosmological constant. Moreover, if the classical spacetime is a homogeneous space $M=G/H$, where $H$ is a certain subgroup of $G$, and the 1-cocycle $\delta$ fulfils the so-called coisotropy condition (see~\cite{BMNphs} and references therein)
\be
\delta(h)\subset h\wedge g,
\qquad
h=\mbox{Lie}(H),
\quad
g=\mbox{Lie}(G),
\label{coisotr}
\ee
then the canonical projection of the Sklyanin bracket to the $M$ submanifold generates a Poisson subalgebra, which is just the Poisson homogeneous space associated to the chosen $r$-matrix, whose quantization will be the quantum homogeneous space associated to $G_z$. 

Finally, it is is also worth mentioning that Lie bialgebras can  alternatively be  described as Drinfel'd double (DD) Lie algebras  given by the  relations
\begin{align}
[X_i,X_j]= C^k_{ij}X_k, \qquad  
[\hat\xi^j,\hat\xi^k]=f^{jk}_i\,\hat\xi^i ,\qquad
[\hat\xi^i,X_j]= C^i_{jk}\hat\xi^k- f^{ik}_j X_k  .\label{DD}
\end{align}
Moreover, 
for any  DD  Lie algebra $D(g)$~\eqref{DD}, its corresponding double Lie group $D(G)$ can be  endowed with a
PL structure generated by the canonical  classical
$r$-matrix
\be
r=\sum_i{x^i\otimes X_i} 
\label{canr}
\ee
which is a solution of the classical Yang--Baxter equation
$[[r,r]]=0$. In fact, if a given even-dimensional Lie algebra can be written in the form~\eqref{DD}, then the quantum deformation associated to the classical $r$-matrix~\eqref{canr} is called a DD one. 

It can be proven that for the (A)dS and Poincar\'e algebras in (2+1) and (3+1) dimensions all 1-cocycles $\delta$ are coboundaries. Therefore, the classification problem for the quantum deformations of these Lie algebras is equivalent to finding all inequivalent (under automorphisms) solutions of the  modified classical Yang--Baxter equation. This is, by no means, a simple task (see~\cite{LBC, Zakr, tallin, LukiBorowiec}). In the sequel we will present the explicit construction and properties of some noncommutative spacetimes with nonvanishing cosmological constant that have been recently obtained through certain quantum  (A)dS deformations of   DD type.


\section{\adsw noncommutative spacetimes from Drinfel'd doubles}

In the specific case of (2+1)-quantum gravity, there are indications~\cite{Starodutsev} that the perturbations of the vacuum state of a Chern--Simons  (CS) quantum gravity theory with cosmological constant $\Lambda$ are invariant under a certain quantum (A)dS algebra, whose zero-curvature limit is the $\kappa$-Poincar\'e quantum algebra~\cite{kappaP}. Also, it is well-known that PL structures on the isometry groups of (2+1) spaces with constant curvature play a relevant role as phase spaces  in  the framework of  a  CS  gauge theory. In this context, the gauge group  is the isometry group,  and the phase space of (2+1)-gravity coupled to point particles is related to the moduli space of flat connections, while the Poisson structure on the moduli space is a PL group (see~\cite{AMII, FR, AT, Witten1, cm1, cm2}). 

Therefore, it seems reasonable to assume that  quantum  deformations of the (A)dS and Poincar\'e groups could play a relevant role in (2+1) quantum gravity. Nevertheless, a plethora of possible quantum deformations for these groups do exist (see~\cite{LBC, Zakr, tallin,LukiBorowiec}), and some physically motivated criteria were needed in order to filter them. An answer to this question was presented in~\cite{BHMplb, BHMCQG}, where it was proven that all the classical $r$-matrices coming from a DD structure of the (A)dS and Poincar\'e groups are compatible with the CS formalism, thus supporting the idea that natural PL  structures for  (2+1)-gravity are classical doubles~\cite{cm2,bernd1}. Moreover, all DD structures for the (A)dS Lie algebras ($so(3,1)$ and $so(2,2)$) were explicitly obtained in~\cite{BHMCQG}, and a similar classification for the Poincar\'e algebra has been also worked out~\cite{BGSHpoinc}. As a result, two main candidates for quantum deformations of the (A)dS symmetries that would be suitable in a (2+1)-gravity CS setting were obtained. The main properties of these two quantum deformations and their associated noncommutative spacetimes will be sketched in what follows. In this approach, the cosmological constant will be included as a `geometric deformation' parameter, and its flat (Poincar\'e) limit can be smoothly obtained.

To this aim we shall make use of the (2+1) \adsw algebra (which can be obtained from (\ref{3mas1}))
\be
\begin{array}{lll} 
[J,P_i]=   \epsilon_{ij}P_j , &\qquad
[J,K_i]=   \epsilon_{ij}K_j , &\qquad  [J,P_0]= 0  , \\[2pt]
[P_i,K_j]=-\delta_{ij}P_0 ,&\qquad [P_0,K_i]=-P_i ,&\qquad
[K_1,K_2]= -J   , \\[2pt]
[P_0,P_i]=\k K_i ,&\qquad [P_1,P_2]= -\k J  ,
\end{array}
\label{21adsw}\nonumber
\ee
where $i,j=1,2$ and  $\epsilon_{12}=1$.  The anti-de Sitter     algebra $so(2,2)$ corresponds to $\k=-\Lambda>0$, the de Sitter   algebra $so(3,1)$ to $\k=-\Lambda<0$ and the Poincar\'e    algebra $iso(2,1)$ to $\k=\Lambda=0$. 

\subsection{A (2+1)  anti-de Sitter  noncommutative spacetime}

The first type of anti-de Sitter classical $r$-matrix coming from a DD structure is given by~\cite{BHMCQG}
\be \label{eq:r1}
r=\eta\, J\wedge K_1  -\tfrac12 \,(-J\wedge P_0  - K_2\wedge P_1 +  K_1\wedge P_2)
\ee
where $\Lambda=-\eta^2=-\k<0$. The associated cocommutator map~\eqref{cocom} reads
\begin{align}\label{cocomm}
&\delta(  J)=-\m     K_2\wedge   J,\qquad \delta(  K_2)=0,\qquad \delta(  K_1)=-\m    K_2\wedge   K_1 , &&\nonumber\\
&\delta(  P_0)= \left(    P_1\wedge  P_2 +\m     P_1\wedge   J-\m^2  K_1\wedge   K_2 \right) ,& &  \\
&\delta(  P_1)= \left(       P_0\wedge   P_2+\m       P_0\wedge   J - \m    P_2\wedge  K_1+\m^2      K_1\wedge   J\right) ,& &\nonumber\\
&\delta(  P_2)= \left(        P_1\wedge  P_0   +\m       P_1\wedge   K_1 -\m^2       J\wedge  K_2\right) .& &\nonumber
\end{align}
From it, the first-order noncommutative spacetime~\eqref{dualL} is defined by the dual algebra $g^\ast$, namely
\be
[\hat x_0,\hat x_1]=\,-\hat x_2, 
\qquad
 [\hat x_0,\hat x_2]=\,\hat x_1,
 \qquad
  [\hat x_1,\hat x_2]=\,\hat x_0,
\label{linear}
\ee
where the $\hat x_i$ coordinates are dual to the translation generators through $\langle \hat x_j,P_k \rangle=\delta_{kj}$. The Lie algebra~\eqref{linear} is just $so(2,1)$, but we have to recall that higher-order contributions can arise from the full quantum coproduct. In fact, the full noncommutative Poisson structure on the homogeneous space $\mathrm{AdS}^{2+1}=\mathrm{SO}(2,2)/\mathrm{SO}(2,1)$ is obtained through the canonical projection of the Sklyanin bracket defined by the $r$-matrix~\eqref{eq:r1}. In terms of the $\mathrm{AdS}^{2+1}$ group coordinates $x_a$  this Poisson noncommutative spacetime reads~\cite{BHMNplb,BMNphs}
\bea
&&   \{x_0,x_1\}  = -\frac{\tanh\m x_2 }{\m} \,\C(x_0,x_1)=- x_2 + o[\m] ,\,  \nonumber\\
&& \{x_0,x_2\} =  \frac{ \tanh\m x_1}{\m}\,\C(x_0,x_1)\, = x_1 + o[\m] ,\label{snyder}\\
&&  \{x_1,x_2\} = \frac{\tan\m x_0}{\m}\,\C(x_0,x_1)\, =  x_0 
+ o[\m] ,
\nonumber
\eea
where $
\C(x_0,x_1)=\cos\m x_0(\cos\m x_0\cosh\m x_1+ \sinh\m x_1)$.
The linearisation of this Poisson bracket  is just the Lie bracket~\eqref{linear}, and higher-order terms turn out to depend on the cosmological constant. Since the 1-cocycle~\eqref{cocomm} fulfils the coisotropy condition~\eqref{coisotr}, then~\eqref{snyder} is a Poisson homogeneous space, although its quantisation  is by no means trivial, as it was discussed in~\cite{BHMNplb}. 

\subsection{A (2+1)    twisted $\kappa$-\adsw noncommutative spacetime}

Another relevant \adsw classical $r$-matrix coming from a different DD structure reads~\cite{BHMCQG,BHMNsigma}
\be
r=z(K_1\wedge P_1+K_2\wedge P_2) +\vt J\wedge P_0 ,
\label{ca}
\ee
where $\vt$ is a second quantum deformation parameter associated to a twist (this is the `time-like' realization of the $r$-matrix, for the `space-like' one we refer to~\cite{BHMNsigma,BMNphs}). Here $\delta$ reads
\begin{eqnarray}
&& \delta(P_0) =  \delta(J)=0 , \nonumber\\ 
&&  \delta (P_1)=   z (P_1\wedge P_0  -\k  K_2\wedge J  ) +   \vt  ( P_0\wedge P_2 +\k  K_1\wedge J)  ,
\nonumber\\ 
&&
  \delta(P_2)=  z  (P_2\wedge P_0+ \k  K_1 \wedge J)     -   \vt   (P_0\wedge P_1 -  \k  K_2 \wedge J)     ,
\nonumber \\
&&
  \delta(K_1)= z (K_1 \wedge P_0  + P_2 \wedge J) +   \vt ( P_0\wedge K_2  -   P_1\wedge J )   ,
 \nonumber\\ 
&&
  \delta(K_2)=   z  (K_2  \wedge P_0 -P_1\wedge J)  -  \vt  (P_0\wedge K_1+P_2\wedge J )  ,
  \nonumber
\end{eqnarray}
from which a first-order noncommutative spacetime is obtained:
\be
[\hat x_0, \hat x_1]=-z \hat x_1 - \vt \hat x_2   ,\qquad
 [\hat x_0, \hat x_2]=-z \hat x_2 + \vt \hat x_1 , \qquad
 [\hat x_1, \hat x_2]=0  .
 \label{cd}\nonumber
\ee
As expected, when the twist parameter $\vt$ vanishes, we recover the well-known $\kappa$-Minkowski spacetime~\cite{kMas,kMR,kZakr,kappaP}. Once again, these are first-order relations, and the cosmological constant  $\Lambda=-\k=-\eta^2$ will appear in higher-order corrections generated by the full quantum deformation. Namely, the full Poisson noncommutative spacetime obtained through the Sklyanin bracket from~\eqref{ca} reads
 \bea  
&& \{x_0,x_1\} =-z\,\frac{\tanh\eta x_1}{\eta \cosh^2\!\eta x_2}- \vt  \cosh\eta x_1 \frac{\tanh\eta x_2}{\eta },\nonumber\\
 &&
\{x_0,x_2\} =-z\,\frac{\tanh\eta x_2}{\eta } + \vt\,  \frac{\sinh\eta x_1}{\eta } ,
\qquad \{x_1,x_2\} =0.
\label{gc}
\eea
The coisotropy condition~\eqref{coisotr} is again fulfilled, and the quantization of~\eqref{gc} can be performed by asuming that $ [\hat x_1,
\hat x_2] =0$. Therefore, the noncommutative twisted $\kappa$-AdS$_\omega$ spacetime  is 
\bea
&&
 [\hat x_0, \hat x_1] =      -z \left( \hat x_1 - \frac 13  \k \hat x_1^3-  \k\hat x_1\hat
x_2^2\right) -    \vt \left(\hat x_2  + \frac 12\k \hat x_1^2\hat x_2 - \frac 13\k   \hat x_2^3 \right) +\mathcal{O}(\k^2) , \nonumber\\ 
&&
[\hat x_0,\hat x_2] =     -z\left(  \hat x_2-\frac 13   \k \hat x_2^3 \right)  +    \vt \left(\hat x_1+ \frac 16 \k \hat x_1^3 \right) +\mathcal{O}(\k^2) ,\label{powerseries}
\nonumber
\eea 
where the cosmological constant generates nonlinear deformation terms that disappear after performing the Minkowski limit $\omega\to 0$.

\subsection{A (3+1) noncommutative \adsw spacetime}

Although there is no CS approach to gravity in (3+1) dimensions, it seems natural to explore whether DD quantum deformations do exist and can be related to the (2+1) ones. The answer is affirmative for the twisted $\kappa$-\adsw deformation, whose DD structure has been recently presented in~\cite{BHNplb, BHMNkappa}. In the basis~\eqref{3mas1}, the corresponding classical $r$-matrix reads
$$
r = z  ( K_1 \wedge P_1 + K_2 \wedge P_2+ K_3 \wedge P_3 + \sqrt{\k}J_1 \wedge J_2  )+\vt  J_3 \wedge P_0 ,
$$
where the term $\sqrt{{\omega}}\, J_{1}\wedge J_{2}$ (which does not exist in~\eqref{ca}) implies that the rotation subalgebra becomes quantum deformed when $\k$ does not vanish. From this expression, it can be shown that  the first-order noncommutative spacetime turns out to be
$$
[\hat x_0,\hat x_1]=- z \hat x_1- \vt   \hat x_2 ,  \qquad  [\hat x_0,\hat x_2]=- z \hat x_2 + \vt \hat x_1, \qquad  [\hat x_0,\hat x_3]=- z \hat x_3, 
$$
together with $ [\hat x_a,\hat x_b]=0 ,\ (a,b = 1,2,3)$. This algebra is not isomorphic to the (3+1) $\kappa$-Minkowski spacetime whenever $\vt$ is not zero (see~\cite{BHNplb}). Again, a nonlinear deformation of this algebra in terms of the cosmological constant is obtained for higher orders~\cite{BGSHpoinc}, thus being the dual counterpart of the full (3+1) Hopf algebra deformation that has been explicitly given in~\cite{BHMNkappa}.

A remarkable problem consists in finding the explicit form of the deformed version of the two \adsw Casimir operators, namely the quadratic one coming from the Killing--Cartan form \be
{\cal C}
= P_0^2-\>P^2  +\k \left(   \>J^2-\>K^2\right) ,
\label{axa}\nonumber
\ee
and the fourth-order invariant
\be{\cal W}
=W^2_0-\>{W}^2+\k \left(\>J\cdot\>K \right)^2,
\label{axb}\nonumber
\ee
where $W_0= \>J\cdot\>P$ and $
 W_a=-  J_a P_0+\epsilon_{abc} K_b P_c$  are the components of the 
Pauli--Lubanski  4-vector. As it has been shown in~\cite{BHMNkappa}, the quantum version of the former is
\bea
{\cal C}_z\!\!\!&=&\!\!\! \frac 2{z^2}\left[ \cosh (zP_0)\cosh(z\sqrt{\k} J_3)-1 \right]+\k \cosh(z P_0) (J_1^2+J_2^2) e^{-z \sqrt{\k}  J_3}\nonumber\\
&& -e^{zP_0} \left( \mathbf{P}^2 +\k \mathbf{K}^2 \right)   \left[ \cosh(z\sqrt{\k}  J_3)+ \frac { z^2\k}2  (J_1^2+J_2^2)  e^{-z \sqrt{\k}  J_3} \right]\label{casa}\\
&&+2\k e^{zP_0}  \left[ \frac{\sinh(z \sqrt{\k}J_3)}{\sqrt{\k}}R_3+ z \left( J_1 R_1 +J_2 R_2+  \frac {z \sqrt{\k}}2 (J_1^2+J_2^2) R_3 \right)  e^{-z \sqrt{\k}  J_3} \right],
\nonumber
\eea
where $R_a=\epsilon_{abc} K_b P_c$.  Indeed, the $\kappa$-Poincar\'e Casimir is obtained as the $\omega\to 0$ limit of (\ref{casa}),
\be
 {\cal C}_z=\frac{2}{z^2}\left[  \cosh(zP_0)-1\right]- e^{z P_0}  \>P^2=\frac{4}{z^2}\sinh^2(zP_0/2) - e^{z P_0}\>P^2.
\label{casdeformeda}
\ee
Therefore, the $\kappa$-Minkowski deformed dispersion relation coming from~\eqref{casdeformeda} turns out to be significantly modified when the cosmological constant does not vanish. This raises the question concerning the physical meaning of the quantum \adsw Casimir~\eqref{casa} in a cosmological context.
Also, the classification of DD structures for the (3+1) (A)dS algebras  is a challenging open problem, and the corresponding Poincar\'e case has been recently solved in~\cite{BGSHpoinc}. Finally, we mention that the characterization of (3+1) (A)dS Poisson homogeneous spaces is currently under investigation by following the approach presented in~\cite{BMNphs}.


\section*{Acknowledgments}

This work has been partially supported by Ministerio de Econom\'{i}a y Competitividad (MINECO, Spain) under grants MTM2013-43820-P and   MTM2016-79639-P (AEI/FEDER, UE), by Junta de Castilla y Le\'on (Spain) under grants BU278U14 and VA057U16 and by the Action MP1405 QSPACE from the European Cooperation in Science and Technology (COST). I.G-S. acknowledges a predoctoral grant from Junta de Castilla y Le\'on and the European Social Fund. 


\end{document}